\documentclass{article}
\usepackage{amsmath,graphicx}

\usepackage{enumitem}
\usepackage{url}
\usepackage{multirow}
\usepackage{booktabs}
\usepackage{subfigure}
\setlist{nosep, leftmargin=14pt}

\usepackage{mwe} 

\usepackage[preprint]{spconf}
\copyrightnotice{\copyright\ IEEE 2000}
\toappear{\parbox{\linewidth}{\textbf{This work has been submitted to the IEEE for possible publication. Copyright may be transferred without notice, after which this version may no longer be accessible.}}}

\title{Tackling Bias in the Dice Similarity Coefficient: Introducing nDSC for White Matter Lesion Segmentation}
\name{\parbox{\linewidth}{\centering
Vatsal Raina$^{1,2}$ \quad Nataliia Molchanova$^{2,3,4}$ \quad Mara Graziani$^{2}$ \quad Andrey Malinin$^{5}$ 
\\ \textit{Henning Muller}$^{2}$ \quad \textit{Meritxell Bach Cuadra}$^{3,4}$ \quad \textit{Mark Gales}$^{1}$}}

\address{
$^{1}$ ALTA Institute, University of Cambridge, UK, \\
$^{2}$ University of Applied Sciences of Western Switzerland (HES-SO), Switzerland, \\
$^{3}$ University of Lausanne, Switzerland, $^{4}$ Lausanne University Hospital, Switzerland, \\
$^{5}$ Shifts Project, Finland.
}
\begin{document}
%
\maketitle
\begin{abstract}
The development of automatic segmentation techniques for medical imaging tasks requires assessment metrics to fairly judge and rank such approaches on benchmarks. The Dice Similarity Coefficient (DSC) is a popular choice for comparing the agreement between the predicted segmentation against a ground-truth mask. However, the DSC metric has been  shown to be biased to the occurrence rate of the positive class in the ground-truth, and hence should be considered in combination with other metrics.
This work describes a detailed analysis of the recently proposed normalised Dice Similarity Coefficient (nDSC) for binary segmentation tasks as an adaptation of DSC which 
scales the precision at a fixed recall rate to tackle this bias. White matter lesion segmentation on magnetic resonance images of multiple sclerosis patients is selected as a case study task to empirically assess the suitability of nDSC. We validate the normalised DSC using two different models across 59 subject scans with a wide range of lesion loads. It is found that the nDSC is less biased than DSC with lesion load on standard white matter lesion segmentation benchmarks measured using standard rank correlation coefficients. 
An implementation of nDSC is made available at: \url{https://github.com/NataliiaMolch/nDSC}.

\end{abstract}
\section{Introduction}
\label{sec:intro}

Semantic segmentation is a common requirement for a diverse range of medical imaging tasks whereby regions must be classified according to a pre-defined, discrete set of options (e.g. lesions, tumours, different types of tissues or structures, etc.).
With increasing demand for medical diagnoses, it is necessary to use automated approaches for segmenting medical images. Such automated approaches must be able to automatically (without human intervention) classify every unit (pixels in 2D images and voxels in 3D images) from a defined set \cite{maier2022metrics}.
To evaluate the quality of medical image segmentation algorithms, it is paramount to have trustworthy evaluation metrics \cite{taha2015metrics}.
Traditionally, the Dice Similarity Coefficient (DSC) \cite{Dice1945MeasuresOT, Srensen1948AMO} (also termed the S{\o}rensen-Dice) is used to measure the quality of the predicted delineations against the manual reference annotations in several subjects, reported by averaging the individual subject-wise DSC scores. 

However, DSC has repeatedly demonstrated to have several weaknesses as an evaluation metric \cite{kofler2021we, reinke2021common, Carass2020EvaluatingWM,maier2022metrics}. Specifically, \cite{maier2022metrics} demonstrated that scans with a larger size of the object to be detected (e.g. lesions or tumours) naturally have a larger DSC score.
We focus on this issue as the DSC has a positive bias with the load of the positive class, such that a subject with a greater presence of the positive class can expect to achieve a greater DSC score for any given segmentation algorithm.
Such a bias makes it difficult to fairly compare the performance of an algorithm on different subjects as it is challenging to disentangle the bias with the segmentation ability (\textit{poor metric selecting} as by \cite{maier2022metrics}). It becomes increasingly important to handle this bias when there is a large variation in the class distributions between different subjects. 
Consequently, medical image segmentation uses a variety of performance metrics to assess the performance of algorithms (cIDice, F$_\beta$ score, IoU \cite{maier2022metrics}) with less emphasis on DSC.

Previous works have proposed assessment metrics such as the Generalised Dice coefficient \cite{crum2006generalized, sudre2017generalised}, which aims to give equal importance to each class when assessing the segmentation quality for a given subject scan when there is an imbalance in the classes to be detected. However, this work is not trying to tackle class imbalance within a given scan but instead tackle the variation in the class distributions \textit{across} multiple scans such that the individual performance on each scan can be fairly compared.
\cite{Carass2020EvaluatingWM} discuss the Refined Sorensen-Dice Analysis metric to look at DSC in more depth for the task of white matter lesion (WML) segmentation. The goal of their work is to avoid drawing conclusions based on the overall DSC and instead study the correlations between DSC and the lesion load. 

In contrast, our goal is to directly improve the DSC and make it unbiased with the positive class. We briefly introduced in \cite{shifts20} the normalized DSC (nDSC) that aims to correct for the bias of DSC for specifically WML segmentation - subjects with a greater fraction of WML voxels are assigned higher DSC scores. Hence, here we will explore the adaptation of DSC to nDSC as a trustworthy performance metric for segmentation algorithms. Our main contribution involves a detailed investigation of the nDSC metric as an improvement to the DSC for medical image binary segmentation tasks. The case of multiple sclerosis lesion segmentation is taken as a proof of concept to show the validity of the method. This application is particularly suited to the analysis as the lesion load can drastically vary across subjects. Let us note that the proposed nDSC metric can also be used in the multi-class segmentation set-up by treating each class to be detected in turn as the positive class, but this extension is beyond the scope of this paper. 


\section{Materials and Methods}



\subsection{Normalised Dice score definition}
\label{sec:definition}

In binary segmentation tasks, each voxel/pixel in an input image must be mutually exclusively classified as either the positive or the negative class. DSC and nDSC assess the similarity between the ground-truth, $Y$, and the prediction, $\hat{Y}$, binary masks by considering counts of true positives TP,
false positives FP
and false negatives FN.
Let $p=\frac{\text{FP}}{\text{TP}}$ and $n=\frac{ \text{FN} }{ \text{TP} }$. 
\begin{align}
\label{eq:metrics1}
\small
\text{DSC} = & 2\left( 2 + p + n \right)^{-1} \\
\small
\label{eq:metrics2}
\text{nDSC} =  & 2\left( 2 + \kappa p + n \right)^{-1}, \hspace{4pt} \kappa = h\left(r^{-1}-1\right)
\end{align}
where,
$h$ represents the ratio between the positive and the negative classes in $\hat{Y}$ while $0<r<1$ denotes a \textit{reference} value that is set to the mean fraction of the positive class in $\hat{Y}$ across a large number of subjects (i.e. the occurrence rate of the positive class averaged across all samples - see  Figure \ref{fig:example_calc}).

\subsection{Rationale}
\label{sec:just}
This section offers an intuitive and theoretically justified explanation of how the adaptation of DSC to nDSC reduces the bias on the rate of occurrence of the positive class. The explanation here will rely on the case study of WML segmentation in magnetic resonance imaging (MRI) scans: a binary segmentation task where each voxel is classified as either background or lesion. Hence, the rate of occurrence is defined as the lesion load - the fraction of voxels in the ground-truth of a subject scan that are attributed to the lesion class. Before understanding how nDSC tackles the bias with lesion load, it is useful to gain intuition on why the bias exists.

Continuing with the notation from Section \ref{sec:definition}, consider the binary ground-truth maps of two separate subject scans, $Y_1$ and $Y_2$. Let $h_1$ and $h_2$ respectively represent the ratio of the lesion voxels to the background voxels in $Y_1$ and $Y_2$. Hence, $h_1/(1+h_1)$ and $h_2/(1+h_2)$ are expressions for the lesion loads in $Y_1$ and $Y_2$ respectively.

The DSC metric (also known as F1) is calculated for each scan using ratios of FP, TP and FN as given in Equation \ref{eq:metrics1}. A separate DSC score, $\text{DSC}_1$ and $\text{DSC}_2$, is calculated for each subject scan using the agreements between the prediction binary masks, $\hat{Y}_1$, $\hat{Y}_2$, with $Y_1$, $Y_2$ respectively. A bias with lesion load means we expect $\text{DSC}_1 > \text{DSC}_2$ if $h_1>h_2$. 

\begin{figure}[htb]
\centerline{\includegraphics[width=7cm]{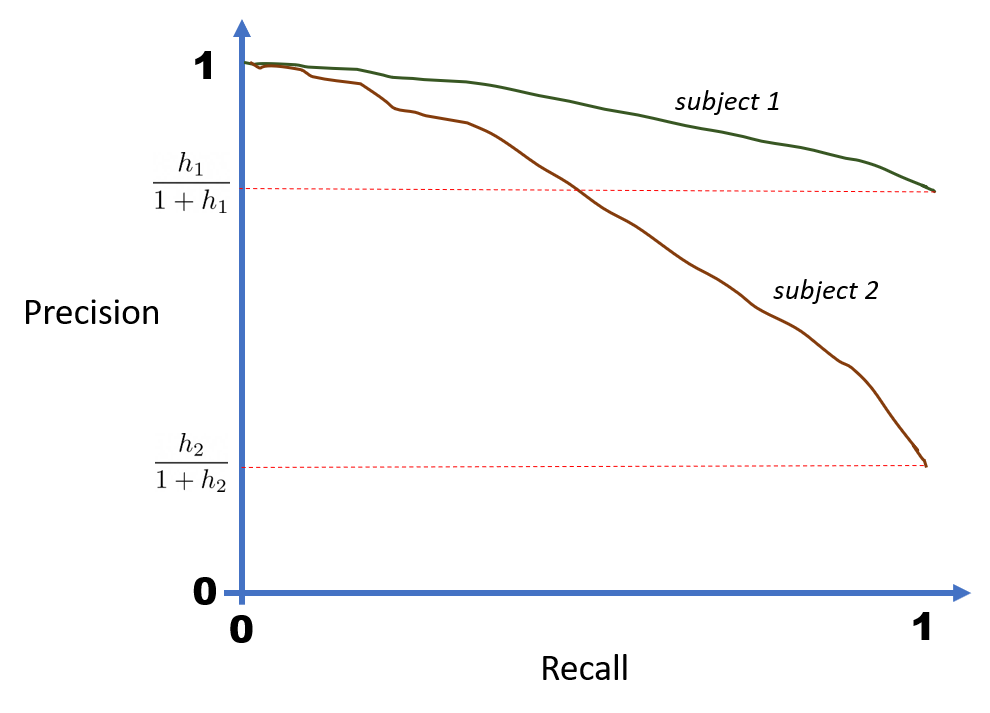}}
\caption{Example precision-recall curves of subject scans.}
\label{fig:example}
\end{figure}

Typically, the prediction binary masks, $\hat{Y}_1$, $\hat{Y}_2$, are derived by thresholding the output probability masks, $\hat{P}_1$, $\hat{P}_2$, where the probability assigned to each voxel is $0\leq p \leq 1$. Therefore, the DSC score can be calculated at different thresholds applied to $\hat{P}$ to generate $\hat{Y}$. A precision-recall curve can capture the performance of a system at all thresholds where precision is the fraction of lesion voxels predicted by the system (at a given threshold) that are correct while recall is the fraction of lesion voxels in the ground-truth that are identified by the system. DSC is the geometric ratio of precision and recall.
\begin{align}
\text{precision} = & (1+p)^{-1} \\
\text{recall} = & (1+n)^{-1} \\
\label{eq:dsc_prec}
\text{DSC} = & 2\times \frac{\text{recall} \times \text{precision}}{\text{recall} + \text{precision}} 
\end{align}
Figure \ref{fig:example} compares typical precision-recall curves for the two subject scans. An ideal precision-recall aims to get as close as possible to the top right corner with coordinate $(1,1)$. Note, at a recall rate of 1.0, the threshold for the probability mask has been set to 0 such that the prediction binary mask assigns all voxels to the lesion class. Therefore, the precision at a recall rate of 1.0 is simply the lesion load of the subject scan. Consequently, the precision-recall curve of subject 1 has to be above the precision-recall curve of subject 2 close to a recall rate of 1.0 if $h_1>h_2$. The precision-recall curve of subject 1 is hence biased to be closer to the ideal shape biasing $\text{DSC}_1 > \text{DSC}_2$ operating at any given recall rate.

The aim of nDSC is to remove the bias at a recall rate of 1.0, which should reduce the bias at other recall rates too.
A strategy to remove the bias can be achieved by scaling the precision at the recall rate of 1.0 to a fixed reference value, $r$. It is sensible to set $r$ to the average lesion load of all the subject scans being considered because such a value ensures the nDSC scores of the scans are in a similar range to the DSC scores. The only variable that can be scaled in precision is the ratio of false and true positives, $p$. Hence, the scaling factor, $\kappa$, for each subject scan can be calculated by the expression given in the latter half of Equation \ref{eq:metrics2} such that $\kappa_1=h_1(r^{-1}-1)$ and $\kappa_2=h_2(r^{-1}-1)$. The calculated value of $\kappa_i$ can then be used to scale the $p$ at the operating recall rate with the expectation that the bias on the lesion load should be reduced.

\begin{figure}[htb]
\centerline{\includegraphics[width=8cm]{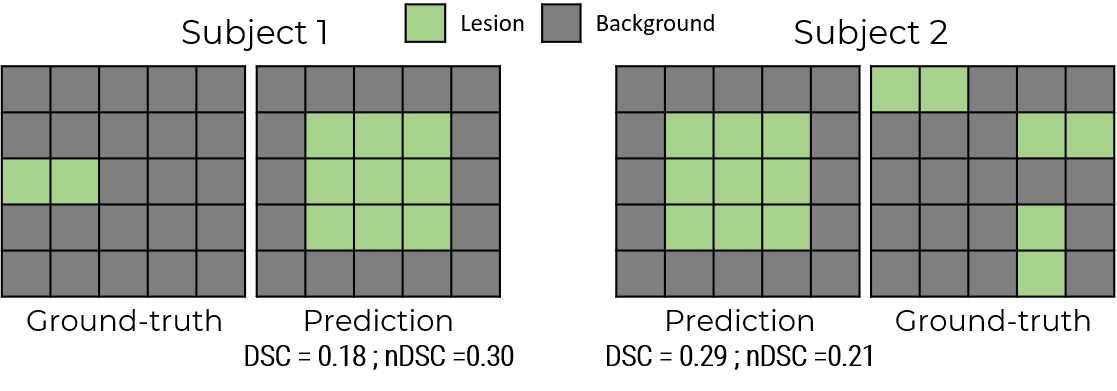}}
\caption{Example illustrative calculation. Reference, $r=\frac{2}{25}$ (average lesion load); $h_1=\frac{2}{23}$, $h_2=\frac{6}{19}$.}
\label{fig:example_calc}
\end{figure}

Figure \ref{fig:example_calc} further illustrates an example calculation of DSC and nDSC for two subjects with different lesion loads. It is clear that DSC suggests the performance on subject 2 is better than subject 1 because of a bias on the lesion load. nDSC re-ranks the performance on the two subjects such that the focus is on segmentation ability instead of rewarding the lesion load in the ground-truth.

\section{Experiments}

\subsection{Set-up}

\begin{figure}[htb]
\centerline{\includegraphics[width=7cm]{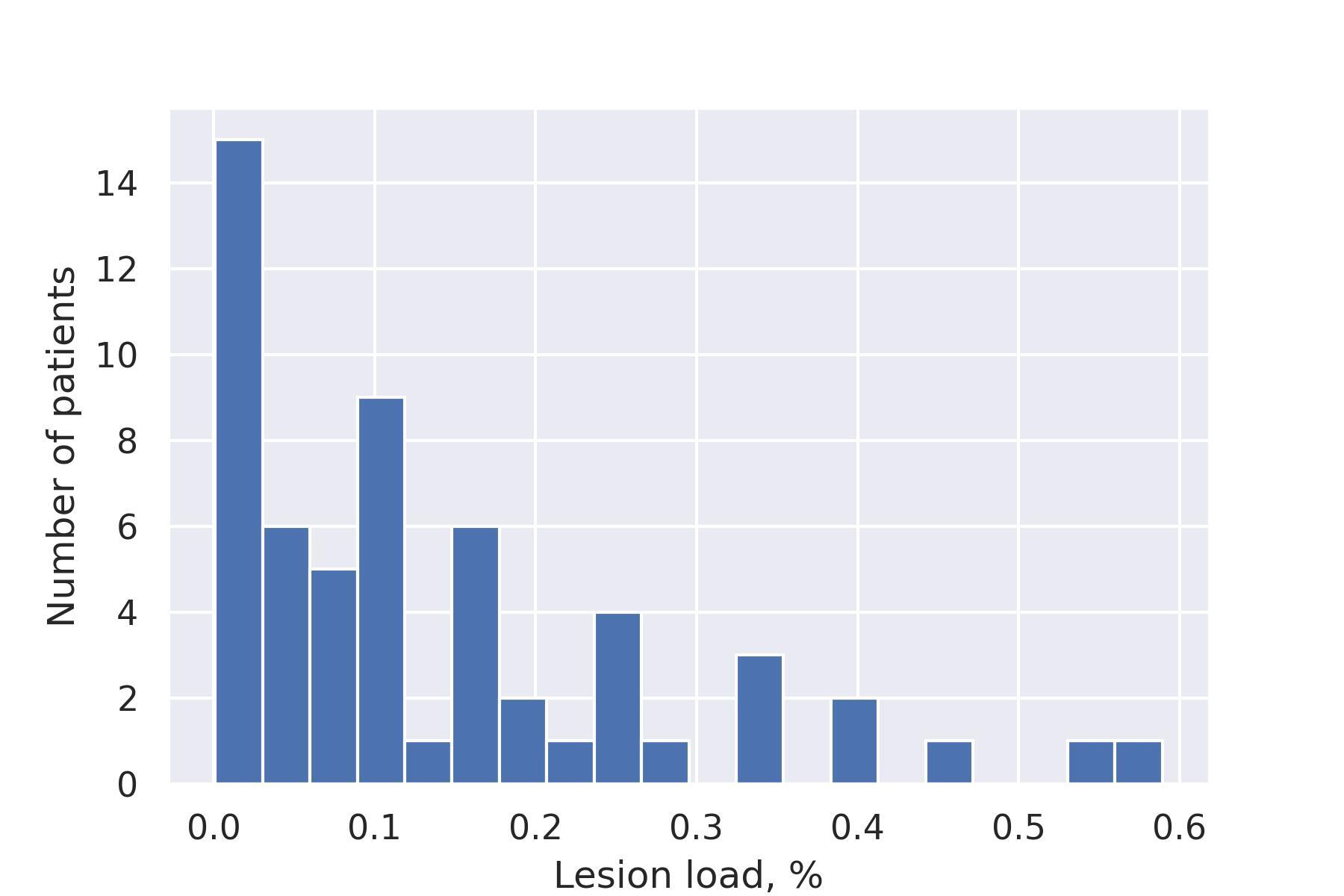}}
\caption{Histogram of lesion loads in the test set.}
\label{fig:hist}
\end{figure}

To empirically assess the bias of DSC and nDSC with the positive class in binary segmentation tasks, a publicly available dataset distributed by the Shifts Project \cite{shifts20} is used for WML segmentation. The dataset is designed for the study of uncertainty estimation across distributionally shifted domains where shifts have been applied across various axes including locations and scanner types but our focus is on comparing the performance metrics used for assessing the performance of the predicted segmentations. The dataset provides FLAIR (fluid-attenuated inversion recovery) scans \cite{flair} which have undergone the preprocessing of denoising, skull stripping, bias field correction and interpolation to a 1mm isovoxel space with corresponding ground-truth binary masks in the same space.
The dataset has been partitioned into \textit{train}, \textit{dev-in}, \textit{eval-in}, \textit{dev-out} and \textit{eval-out} where \textit{dev-in} and \textit{eval-in} are in-domain with the training data while \textit{dev-out} and \textit{eval-out} are distributionally shifted with respect to \textit{train}. In the experiments for this paper, we maintain these splits but combine the publicly available test splits into a single evaluation set. Hence, our segmentation models are trained using the 33 scans from \textit{train} with hyperparameter tuning on 7 scans from \textit{dev-in} and evaluation on \textit{eval-in} + \textit{dev-out} such that the test set consists of 59 images. Note, \textit{eval-out} is not publicly available as it is used for official benchmarking of systems for the Shifts Project and hence is excluded in this work. 
Figure \ref{fig:hist} presents the distribution of the lesion loads in the test set. The order of magnitude variation in lesion load from less than 0.02\% to greater than 0.5\% allows the bias of each performance metric to be assessed with the lesion load.
The test set is additionally partitioned into two subsets: Test = TestA + TestB. TestA consists of 30 scans with the lowest lesion loads of $4.07\pm 3.33 $\% to one standard deviation while TestB consists of 29 scans with the highest lesion loads of $24.70\pm 13.26$\%. This synthetic partition allows an explicit assessment of the bias of a performance metric with the lesion loads.

Two model architectures are considered. The first is based on a shallow 3D U-net with a depth of 3 that was used in \cite{shifts20} as a baseline and additionally proven to be competitive for WML segmentation \cite{LAROSA2020102335}.
Second, UNETR \cite{unetr2021} is considered which has a transformer-based encoder with a convolutional decoder.
Both models were trained for a maximum of 300 epochs with early-stopping. A loss function (combination of Dice and focal loss) was optimised using an Adam algorithm with a constant learning rate of 1e-5.
The architectures take as an input $96\times96\times96$ sub-volumes of FLAIR scans, which are aggregated using Gaussian-weighted averaging at inference time.
The final system is a deep ensemble of 3 individual models, each trained with a different initialisation.
Binary segmentation masks are obtained by thresholding the probability maps averaged across different seeds at 0.35, which was chosen on the validation set by optimising the nDSC/DSC.

The bias of a given metric (DSC or nDSC) with the lesion load is assessed using the Spearman's rank correlation coefficient, $-1\leq \rho \leq 1$. For the test data, $\rho$ is calculated by comparing the ranks of each subject scan according to the performance metric and the lesion load where $\rho$ close to 1 suggests there is a strong bias with lesion load, $\rho$ approaching -1 indicates a strong negative bias while $\rho=0$ means the metric is unbiased with the lesion load.
Additionally, Kendall's tau, $\tau$, is used to measure the correspondence between the two sets of rankings where $-1\leq \tau \leq 1$.
For these experiments, $r$ in Equation \ref{eq:metrics2} is set to 0.001 as this is roughly the average lesion load across the \textit{train} split subject scans.

    

\subsection{Results}

\begin{table}[htbp!]
\small
\centering
    \begin{tabular}{l|ccc|ccc}
    \toprule
    \multirow{2}*{Model} & \multicolumn{3}{c|}{DSC} & \multicolumn{3}{c}{nDSC}  \\
     & Test & TestA & TestB & Test & TestA & TestB \\
    \midrule
UNET & 0.574 & 0.502 & 0.652 & 0.601 & 0.584 & 0.619 \\
UNETR & 0.632 & 0.542 & 0.729 & 0.646 & 0.615 & 0.678 \\
   \bottomrule
    \end{tabular}
\caption{Performance between metrics and the lesion load for the UNET and UNETR model ensembles. Test A and B are datasets with average low, and high lesion load, respectively.}
    \label{tab:performance}
\end{table}

Table \ref{tab:performance} presents the performance of WML segmentation using UNET and UNETR according to the averaged DSC and nDSC scores across all the scans. UNETR model out-performs UNET according to both the DSC and nDSC metrics.
Using DSC as a performance metric suggests that both models are substantially better on TestB compared to TestA. This is a misleading conclusion as the higher performance on TestB is due to the larger lesion load in the subject scans. In contrast, nDSC is less biased to lesion load as observed with a smaller gain in its value between TestA and TestB.

\begin{table}[htbp!]
\small
\centering
    \begin{tabular}{l|cc|cc}
    \toprule
    \multirow{2}*{Model} & \multicolumn{2}{c|}{$\rho$} & \multicolumn{2}{c}{$\tau$}  \\
     & DSC & nDSC & DSC & nDSC \\
    \midrule
UNET & 0.481 & 0.056 & 0.365 & 0.033 \\
UNETR & 0.582 & 0.106 & 0.431 & 0.067 \\

   \bottomrule
    \end{tabular}
\caption{Correlation coefficients between metrics and the lesion load for the UNET and UNETR ensembles on Test.}
    \label{tab:correlation}
\end{table}

Table \ref{tab:correlation} gives rank correlation coefficients (Spearman's and Kendall's tau) between the individual DSC/nDSC scores and the lesion loads of each subject scan for the full test set using both the UNET and the UNETR. Evidently, DSC is more biased towards the lesion load than nDSC for both models as the correlation coefficients with nDSC are a lot closer to the unbiased value of 0.0.

\begin{figure}[htbp!]
\centerline{\includegraphics[width=7cm]{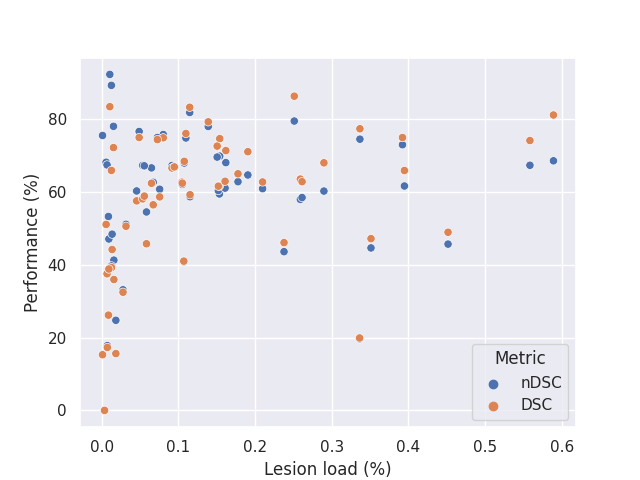}}
 \caption{Correlation with lesion load using UNET.}
 \label{fig:ndsc_correlations}
\end{figure}

Figure \ref{fig:ndsc_correlations} visualises how the absolute values of DSC and nDSC vary with the lesion load for the UNET. It is apparent that DSC has a clear correlation with the lesion load, reflected by $\rho=0.481$, $\tau=0.365$ while nDSC flattens the correlation to reduce the bias with lesion load. nDSC prevents the subject scans with low lesion loads being unfairly penalised while equally avoids over-rewarding the scans with high lesion loads.

\begin{figure}[htb]
\centerline{\includegraphics[width=7cm]{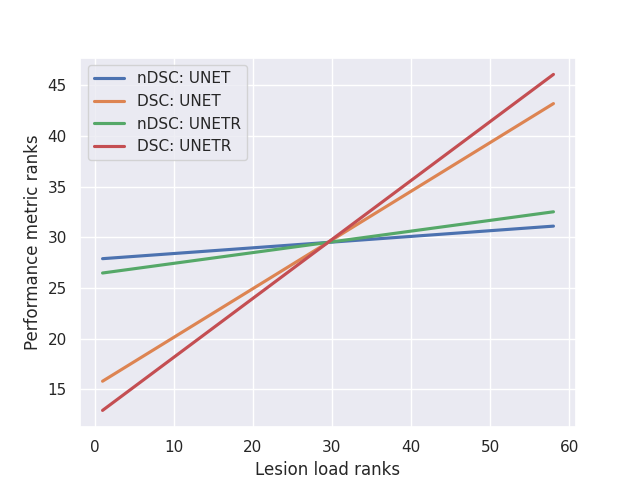}}
\caption{Ranks of the metrics with the lesion load.}
\label{fig:reg_bias}
\end{figure}

Finally, Figure \ref{fig:reg_bias} plots a regression line of the ranks  for both the performance metrics, DSC and nDSC, against the lesion load for the UNET and UNETR models. A gradient closer to 1 implies perfect correlation and hence bias between the  metric and the lesion load. It is clear from the graph that the nDSC regression lines are flatter than the DSC metric regression lines for both the models, indicating empirically that nDSC is less biased than the DSC performance metric relative to the lesion load for the task of WML segmentation.

\section{Conclusions}
We thoroughly examine theoretically and empirically the new nDSC metric in relation to the popular DSC metric using the binary segmentation task of white matter lesion segmentation from brain scans as a case study. nDSC is a favourable performance metric over DSC in situations where there is large variation in the distribution of the segmentation classes across different individuals of a cohort study.
Future work should investigate the applicability of the nDSC metric for other popular medical image segmentation tasks including tumor and fetal brain tissue segmentation tasks. For the latter, the development of a generalised nDSC metric for the multi-classification segmentation set-up would be needed. 

\section{Compliance with ethical standards}
\label{sec:ethics}

This research study was conducted retrospectively using human subject data made available in open access by \cite{shifts20}. Ethical approval was not required as confirmed by the license attached with the open access data.

\section{Acknowledgments}
\label{sec:acknowledgments}
This work was supported by the Hasler Foundation Responsible AI programme (MSxplain) and the EU Horizon 2020 project AI4Media (grant 951911).
The research is further supported by the EPSRC (The Engineering and Physical Sciences Research Council) Doctoral Training Partnership (DTP) PhD studentship.
We acknowledge access to the facilities and expertise of the CIBM Center for Biomedical Imaging, a Swiss research center of excellence founded and supported by Lausanne University Hospital (CHUV), University of Lausanne (UNIL), École polytechnique fédérale de Lausanne (EPFL), University of Geneva (UNIGE) and Geneva University Hospitals (HUG).

\bibliographystyle{IEEEbib}
\bibliography{strings,refs}

\end{document}